\documentclass[%
 reprint,
 amsmath,amssymb,
 aps,
 pre,
]{revtex4-1}

\usepackage{graphicx}
\usepackage{dcolumn}
\usepackage{bm}
\usepackage{amsmath} 
\usepackage{ amssymb} 
\usepackage{amsfonts}
\usepackage{txfonts}

\begin{document}

\preprint{APS/123-QED}
\title{Manifold Learning Approach for Chaos in the Dripping Faucet}

\author{Hiromichi Suetani$^{(a, b, c)}$, 
 Karin Soejima$^{(a)}$, Rei Matsuoka$^{(d)}$ Ulrich Parlitz$^{(e, f)}$ and Hiroki Hata$^{(a)}$}
 \affiliation{%
${}^{(a)}$Department of Physics and Astronomy, Kagoshima University, Kagoshima 890--0065,  Japan
}
 \affiliation{%
${}^{(b)}$Decoding and Controlling Brain Information, PRESTO, JST,  Saitama 332--0012,  Japan
}
 \affiliation{%
${}^{(c)}$Flucto-Order Functions Research Team, RIKEN--HYU Collaboration Research Center, RIKEN Advanced Science Institute, Saitama 351--0198,  Japan
  } 
   \affiliation{%
${}^{(d)}$Department of Energy Engineering and Science, Nagoya University, Nagoya 464--8603 Japan
  } 
   \affiliation{%
${}^{(e)}$Biomedical Physics Group, Max Planck Institute for Dynamics and Self-Organization, 
Am Fa\ss berg 17, 37077 G\"ottingen, Germany
  } 
     \affiliation{%
${}^{(f)}$Institute for Nonlinear Dynamics, Georg-August-Universit\"at G\"ottingen,
Am Fa\ss berg 17, 37077 G\"ottingen, Germany
  } 



\date{\today}

\begin{abstract}
Dripping water from a faucet is a typical example exhibiting rich nonlinear phenomena.
For such a system, the time stamps at which water drops separate from the faucet can be directly observed in real experiments, and the time series of intervals $\tau_{n}$ between drop separations becomes a subject of analysis.   
Even if the mass $m_n$ of a drop at the onset of the $n$-th separation, which cannot be observed directly, exhibits perfectly deterministic dynamics, it sometimes fails to obtain important information from time series of $\tau_n$.
This is because the return plot $\tau_{n-1}$ vs. $\tau_{n}$ may become a multi-valued function, i.e., not a deterministic dynamical system. 
In this paper, we propose a method to construct a nonlinear coordinate which provides a ``surrogate" of the internal state $m_n$ from the time series of $\tau_n$.
Here, a key of the proposed approach is to use ISOMAP, which is a well-known method of manifold learning. 
We first apply it to the time series of $\tau_n$ generated from the numerical simulation of a phenomenological mass-spring model for the dripping faucet system. 
It is shown that a clear one-dimensional map is obtained by the proposed approach, whose characteristic quantities such as the Lyapunov exponent, the topological entropy, and the time correlation function coincide with the original dripping faucet system.
Furthermore, we also analyze data obtained from real dripping faucet experiments which also provides promising results.     
 
\begin{description}
\item[PACS numbers] 05.45.-a, 05.45.Tp, 05.10.-a
\end{description}
\end{abstract}

\maketitle


\section{\label{sec:level1}Introduction}
Dripping of water from a faucet is ordinarily seen in our daily life. 
At first glance, such a motion of dripping looks very common.
It provides, however,  a variety of rich nonlinear dynamics including the period-doubling bifurcation to chaos, intermittency, crisis, hysteresis, and etc.
In particular, Robert Shaw and his collaborators~\cite{Shaw84} first found that there is a clear transition from a periodic motion to low-dimensional chaos by investigating the time intervals $\tau_n$ between dripping separations from the faucet,  both theoretically and experimentally.

The dripping water is fluid dynamics, i.e., ideally described as an infinite-dimensional dynamical system.
But as far as the dynamics is confined within a low dimensional attractor,  it can be modeled by a  class of phenomenological models called  ``mass-spring" systems~\cite{Shaw84}.
Since the pioneering work of Shaw et al., many versions of the mass-spring system for the dripping faucet have been proposed.  
Among them, Kiyono and Fuchikami~\cite{Kiyo99} significantly improved  the mass-spring system on the basis of both,  numerical simulations of fluid dynamics~\cite{Fuch99}  and real experiments~\cite{Kats99}.
They showed that their model 
can systematically explain various aspects of the complex behaviors observed in the real dripping faucet experiments.  
 
One of the most prominent aspects of  the Kiyono--Fuchikami model is that the essential feature of chaos in the dripping faucet is exactly represented as a one-dimensional map. 
More precisely, the mass $m_n$ at the moment of the $n$--th separation of a drop from the faucet obeys a one-dimensional mapping dynamical system, i.e., there exists a deterministic scalar function $f(\cdot)$ such that  $m_{n} = f(m_{n-1})$.

In general, however, not all state variables are observable in real experiments.
In the case of the dripping  faucet system, for example, it is very difficult to observe the mass $m_n$ of a drop in a direct way.
Instead, time intervals $\tau_n$ between successive drop separations can be recorded in real experiments. 
As investigated by Shaw et al., depending on the degree of flux of water, the return plot $\tau_{n-1}$ vs. $\tau_{n}$ also shows a  clear functional relationship.
At the same time, however, they have also shown that it often takes the form of a multi-valued relation.
Namely, there are two or more candidates of $\tau_n$ against a single value of $\tau_{n-1}$, which prevents us from interpreting the dripping faucet as a simple one-dimensional mapping system.
This multi-valuedness problem often occurs in general chaotic dynamical systems such as the Kuramoto-Sivashinsky equation~\cite{Chr97}. 

On the other hand, the existence of the one-dimensional map $f$ associated with the mass $m_n$ means that an embedding of the dripping-time interval $\tau_n$ into a sufficiently high, say $d$--dimensional Euclidean space $\mathbb{R}^d$ as $\boldsymbol{s}_n = (\tau_{n-d+1}, ..., \tau_{n-1}, \tau_n)$, is lying on a one-dimensional manifold ${\cal S} \subset \mathbb{R}^d$.
Then, a point $\boldsymbol{s}_n \in {\cal S}$ obeys a deterministic law as $\boldsymbol{s}_{n}=\mathbb{F}(\boldsymbol{s}_{n-1} )$ where $\mathbb{F}(\cdot)$ is a $d$-dimensional vector valued function whereas the relationship between $\tau_{n-1}$ and $\tau_{n}$ is a multi-valued one.
Actually, in the case of the Kiyono and Fuchikami's mass-spring model,  embedding $\tau_n$ into a three-dimensional space generally results a filament-like one-dimensional manifold without crossing. 
Therefore, if a new coordinate $u$ is spanned along ${\cal T}$, which plays the role of a surrogate variable for the internal state $m_n$, then we obtain a more simplified expression as $u_{n} = \varg (u_{n-1})$ where $\varg (\cdot)$ is a  scalar function of $u$. 
Even if the original one-dimensional mapping system $m_{n} = f(m_{n-1})$ is not available, important dynamical features can be obtained from the mapping associated with the surrogate variable $u$ as  $u_{n} = \varg (u_{n-1})$. 

To identify lower dimensional representations of the dynamics, dimension reduction methods can be employed.
Dimension reduction is an important task of data \mbox{(pre-)}  processing with applications in pattern and speech recognition, image processing, bioinformatics, psychology, etc. 
Linear subspaces containing or approximating the available data can be identified using Principal Component Analysis (PCA), Independent Component Analysis (ICA), and many useful methods in various fields~\cite{HTF09}.
However, when the data set of interest is located on or close to a (sub-) manifold with significant curvature, the applicability of these linear methods is limited and nonlinear dimension reduction methods have to be employed.
%

Recently, in the field of statistical  machine learning, methods of {\it manifold learning} have been developed for providing a low-dimensional representation when data is lying on  a nonlinear low dimensional manifold embedded in a high dimensional  Euclidean space.
A number of methods have been proposed, and in the present study, we employ ISOMAP~\cite{Tene00} for such a purpose.
ISOMAP, which is an abbreviation of the term ``isometric feature mapping",  is a method of manifold learning where the geodesics between training samples are employed as the dissimilarity information in multi-dimensional scaling (MDS)~\cite{Cox00}.  

In this paper, we demonstrate that ISOMAP is very useful to extract a surrogate state variable $u$ and to construct a well defined one-dimensional map $\varg(\cdot)$ for both, numerical and real experimental data. 
It is shown that dynamical characteristics such as the Lyapunov exponent and the time correlation function can be computed from $\varg(\cdot)$.    

The present paper is organized as follows.
In Section II, we explain the dripping faucet system.
In Section III, we first introduce the method of ISOMAP, then we apply it to data generated from the  mass--spring model mentioned in the previous section.
Finally, in Section IV, we give a summary and discuss possible directions of future research.

\section{\label{sec:level2}Dripping Faucet System: Model and Experiment}
\subsection{Basic mechanism}
\begin{figure}
\begin{center}
\includegraphics[width=8cm,clip]{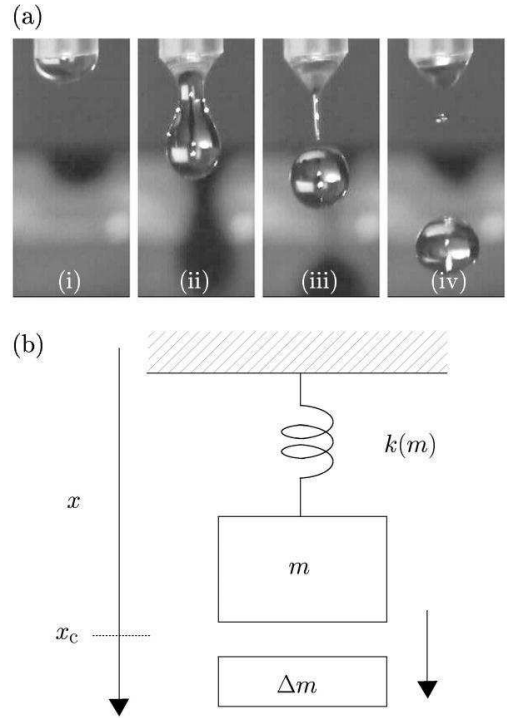}
\\
\caption{(a) Snapshots of a high speed movie showing the separation of a drop. (b) Illustration of the mass spring model.}
\label{f_s1}
\end{center}
\end{figure}

Let us begin with a brief introduction of the basic mechanism how a water drop separates from a faucet.
Figure \ref{f_s1} (a) shows a snapshot at just the moment when a water drop separates from a burette  in experiments.
Here, the shape of the drop is determined by the balance between the surface tension and the weight of water.
When increasing the mass of a drop by injecting water, the following processes are repeated with time.
(i)  A ``neck" which connects between the drop and the faucet is formed by the break of the balance between the tension and the mass of water.
(ii) When the weight reaches a critical value, the neck is broken, i.e., a portion of the drop separates from the faucet. 
(iii) Just after its separation, the remainder of the drop rapidly shrinks by the surface tension to the upward direction. 
(iv) Finally, the drop grows again with oscillations.      
\subsection{Mass-spring model}
Based on observations as mentioned in the previous subsection,
the following equations of motion can be considered as a phenomenological model for the dripping faucet experiment~\cite{Shaw84} 
\begin{eqnarray}
\label{eq2-1a}
&& m\ddot{x} + \dot{x}\dot{m} =  - kx -\gamma x + mg, \\
\label{eq2-1b}
&& \dot{m}  =  Q \ ({\rm const.}).
\end{eqnarray}
Such a model is called a mass-spring model and its schematic illustration is depicted in Fig. \ref{f_s1} (b).
Here, $x$ is the vertical position of the forming drop to the downward direction, $m$ is its mass, $g$ is the gravitational acceleration, $k$ is the stiffness of the spring, and $\gamma$ is the damping parameter.
The restoring force given by the surface tension is represented as a spring force in Eq.~(\ref{eq2-1a}), and the mass of the forming drop linearly depends on time with the rate $Q$ as described in Eq.~(\ref{eq2-1b}), because a drop grows with time due to the influx of water from the faucet.
It is also assumed that 
when the position of the drop reaches the critical point $x_c$,  the drop loses its mass by $\Delta m$ due to
the separation of a portion of the drop and this portion falls into the ground.
In spite of its simplicity, this model can explain many dynamical aspects of the real dripping faucet~\cite{Shaw84}.

Kiyono and Fuchikami improved the above phenomenological model~\cite{Kiyo99}  based on the knowledge of the numerical simulations of fluid dynamicsi~\cite{Fuch99}  and real experiments of the dripping faucet~\cite{Kats99}.
They first modified  the equation of motion in Eq.~(\ref{eq2-1a}) as 
\begin{eqnarray}
m\ddot{x} + (\dot{x}-v_0)\dot{m} =  - kx -\gamma x + mg,
\label{eq2-1c}
\end{eqnarray}
where $v_0$ is the velocity of the influx of water.
Note here that there is a relation between $v_0$ and $Q$ in Eq.~(\ref{eq2-1b}) as $Q=\pi a^2 v_0$ where $a$ is the radius of the faucet.
Then, based on their real experiments, they considered that the stiffness $k$ in Eq.~(\ref{eq2-1c}) also depends on the mass of the drop as: 
\begin{eqnarray}
\label{eq2-2}
k(m) = 
\left \{
\begin{array}{ll}
-11.4m + 52.5 \quad (m<m_c), \\
0 \quad (m\geq m_c), 
\end{array}
\right .
\end{eqnarray}
\\
where $m_c=4.61$.
Equation (\ref{eq2-2}) means that when the mass $m$ amounts to $m_c$, the value of the stiffness becomes zero, then the drop undergoes free-fall.
In their experiments, the units of the length, time, and mass are chosen as $(\gamma/\rho g)^{1/2}$(=0.27cm), $(\Gamma/\rho g^3)^{1/4}$(=0.017 sec) and $\rho (\gamma/\rho g)^{3/2}$=(0.020$g$), respectively, where $\Gamma$ is the surface tension and $\rho$ is the density.
Using these units, parameters are set to $\gamma=0.05$, $g=1$, $x_c = 5.5$, $\Delta m = 0.8 m - 0.3$, and $a=0.916$, and the constants in Eq.~(\ref{eq2-2}) are also determined from their experiments.
They also assumed that just after a portion of the drop separates, the position and velocity are reset to  $x=x_0 = 2.0, \ \dot{x}=\dot{x}_0 = 0$.

\begin{figure}
\begin{center}
\includegraphics[width=7cm,clip]{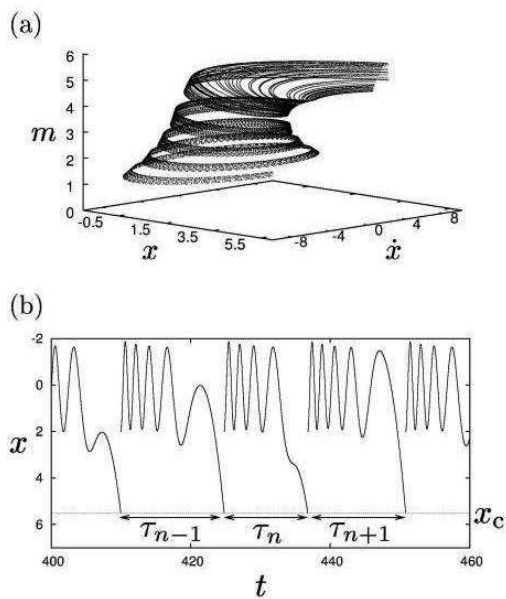}
\caption{(a) Chaotic trajectory of the Kiyono-Fuchikami model. (b) Time series of drop position $x$ and the time interval $\tau$ between drop separations.}
\label{f_s2}
\end{center}
\end{figure}

Figure \ref{f_s2} shows a trajectory of the above mentioned model with $v_0=0.1130$  after some transient.
In Fig. \ref{f_s2} (a) we can see that the trajectory is tracing a chaotic attractor.
In real experiments, however, it is in general impossible to observe all state variables of the system.
In the case of the dripping faucet, time intervals $\tau_n, (n=1,2,...)$ between successive drop separations are observed in experiments.
How to determine $\tau_n$ from the signal of the position $x(t)$ is depicted in Fig. \ref{f_s2} (b). 
Here the variable $m_n$ is the value of the mass at the moment of the $n$-th drop separation.  

\begin{figure}
\begin{center}
\includegraphics[width=8.5cm,clip]{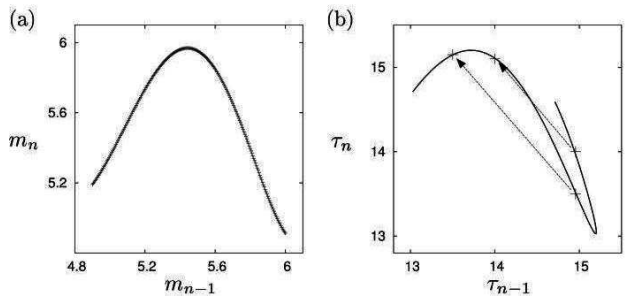}
\\
\caption{(a) Return plot of $m_{n-1}$ vs. $m_n$. (b) Return plot of $\tau_{n-1}$ vs. $\tau_n$. Movements of points as the time delay embedding vector ${\boldsymbol s}_n = (\tau_{n-1}, \tau_{n})$ are also depicted. }
\label{f_s3}
\end{center}
\end{figure}

Figure \ref{f_s3} (a) shows the return plot $m_{n-1}$ vs.  $m_{n}$ for $v_0=0.1130$. 
We can see that there is a clear scalar function between $m_{n-1}$ and  $m_{n}$ as $m_n = f(m_{n-1})$. 
In Figure \ref{f_s3} (b), however, the return plot $\tau_{n-1}$ vs.  $\tau_{n}$ is a multi-valued function, i.e., the right-hand side of the return plot shows the 1 to 2 values. 
From a different viewpoint, if we regard this return plot as the time-delay embedding of $\tau_n$ into the two-dimensional plane  ${\mathbb R}^2$ as ${\boldsymbol s}_n = (\tau_{n-1}, \tau_{n})$ denoting the manifold on which the states ${\boldsymbol s}_n$ are lying as ${\cal S}$, 
then, we can see that there is a deterministic relationship between ${\boldsymbol s}_{n-1}$ and ${\boldsymbol s}_{n}$ as ${\boldsymbol s}_{n} = {\mathbb F} ({\boldsymbol s}_{n-1})$  in ${\mathbb R}^2$.
\subsection{Bifurcation Structure}
\begin{figure}
\begin{center}
\includegraphics[width=8cm,clip]{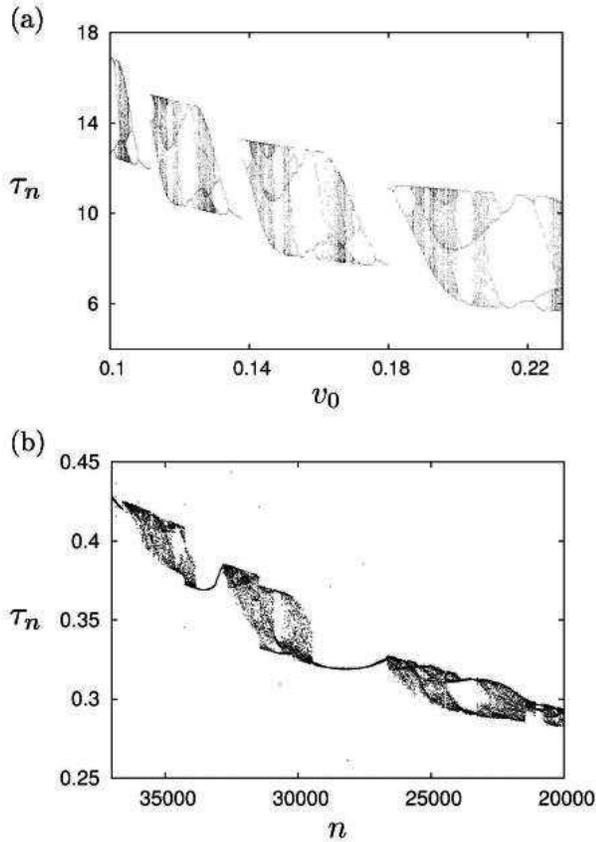}
\caption{(a) Bifurcation diagram of the Kiyono-Fuchikami model. Note that the unit of time is rescaled, e.g., $\tau = 14$ corresponds to 0.238 sec. (b) Time series of a real dripping faucet experiment.}
\label{f_s4}
\end{center}
\end{figure}

We also investigated how the statistical property of the mass-spring model (Eqs.~(\ref{eq2-1c}) and (\ref{eq2-2})) depends on the water influx $v_0$, and the result is shown in Fig.~\ref{f_s4} (a).
One can see repetitions of the period doubling bifurcation route to chaos, as well as periodic windows and  their reverses as increasing $v_0$. 
 We also made experiments to check whether the true dripping faucet system also exhibits this bifurcation structure. 
Figure \ref{f_s4} (b) shows a time series of the time intervals of drop separations
 over a long time period. 
Here, in our experiments, the surface of water of the bath decreases very slowly because no water is supplied from outside,  which plays a role of changing the water influx.
Therefore, this figure represents a kind of ``bifurcation" diagram.     
One can see that there is a significant qualitative similarity between the numerical simulations (Fig.~\ref{f_s4} (a)) and the real experiments (Fig.~\ref{f_s4} (b)).  

\section{Extracting one-dimensional maps of internal state variables by ISOMAP}
As shown in the previous section, this dripping model is essentially described by a one-dimensional map $f(m)$.
Its experimental observables $\tau_n$ also show a one-dimensional filament in the $d$-dimensional space $(\tau_{n-d+1},\tau_{n-d+2},\dots,\tau_n)$, but the relationship between $\tau_{n-1}$ vs. $\tau_{n}$ is not always given by a one-dimensional map directly.
In this paper, we discuss the case $d=2$ as shown in Fig.\ref{f_s3}.
This suggests that the dripping-time interval $\tau$ isn't appropriate for the simple description of the dynamics.
If we can construct a new coordinate $u$ along the filament, the dynamics must be described by a one-dimensional map $u_{n}=\varg(u_{n-1})$ and easily analyzed using the theory for one-dimensional maps.  
In this section, we try to construct a new coordinate $u$ 
by applying ISOMAP to the time series $\tau_n$ and get the one-dimensional map $\varg(u)$.
In addition, we test whether we can recover the statistical properties of the dripping faucet system from the one-dimensional map $\varg(u)$.
\subsection{ISOMAP}
ISOMAP is one of several widely used low-dimensional embedding methods, which is an extension of classical MDS (multi-dimensional scaling) \cite{Tene00}.
MDS seeks a low dimensional representation of the sample points.
This is achieved by plotting data points in a low dimensional space preserving the "dissimilarity" (generalized distance) between sample points (in the original higher dimensional space) as much as possible. 
For the dripping time series $\tau_n$, when we use the Euclidian distances (pairwise distances between sample points) in the
($\tau_{n-1}, \tau_n$) plane as the dissimilarity, we shall not obtain a one-dimensional embedding, because the sample points are located on a curved filament.
In ISOMAP, the geodesic distance on the low dimensional structure instead of the Euclidean distance is used and then we can embed 
the sample points
into the one-dimensional space and get the new coordinate $u$.
A concrete procedure is described as follows. 

\begin{figure}
\begin{center}
\includegraphics[width=8.5cm,clip]{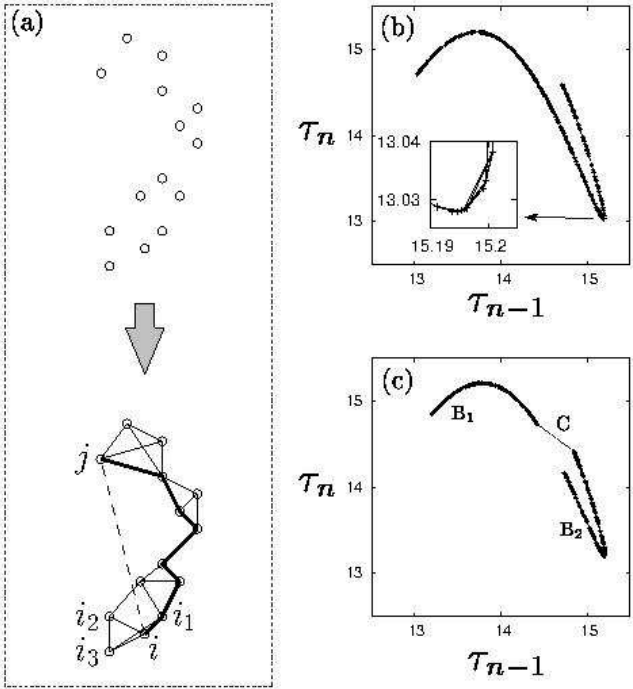}
\end{center}
\caption{
Neighboring graph $\cal G$.
(a) Schematic illustration of the sample points ($\circ$) and the 3-neighboring graph $\cal G$ (solid line). The heavy solid line is the shortest path between points $i$ and $j$ and the dashed line is the Euclidean distance. (b) 500 sample points (+) and 4-neighboring graph $\cal G$ (line) for $v_0 = 0.1130$.
(c) Graph $\cal G$ for 2-band chaos ($v_0=0.1128$). The two clusters B$_{1}$, B$_{2}$ caused by 2-band chaos
are connected by the shortest path C.
}
\label{f_hh2}
\end{figure}

First, we compute the geodesic distance $d_{ij}$ between $i$-th and $j$-th sample points which is approximated with the shortest path from one to the other on the neighboring graph $\cal G$.
This graph $\cal G$ is constructed by locally connecting among sample points  (we employ the Euclidean distance to construct $k$-nearest neighbors). 
This procedure with $k=3$ is illustrated in Fig. \ref{f_hh2}(a). The point $i$ is connected to $i_1,i_2,i_3$ which are $k$-nearest neighbors of the point $i$ and all sample points are connected in the same manner.
The distance $d_{ij}$ between points $i$ and $j$ is not defined by the Euclidean distance  (dashed line), but by the shortest path distance (heavy solid line).
As a result, $d_{ij}$ is an approximation of the geodesic distance on the manifold.
Figure \ref{f_hh2}(b) shows the graph $\cal G$ of the dripping faucet (\ref{eq2-1c}) and (\ref{eq2-2}) for  $v_0 = 0.1130, N = 500$ and $k = 4$.
And $d_{ij}$ is obviously a good approximation of the geodesic distance along the filament.

The bifurcation diagram for this dripping faucet is shown in Fig. \ref{f_s4}. 
When the attractor has $n$-bands or exhibits strong intermittency, the neighboring graph $\cal G$ may be separated into a few clusters and then we cannot estimate the geodesic distances $d_{ij}$.
In this case we adopt the shortest connection among the clusters to construct the global graph $\cal G$.
A example for two-band chaos ($v_0=0.1128$) is shown in Fig \ref{f_hh2}(c).

The second step is MDS whose purpose is to place a set of new points $\
{u}_i, (i=1,2, \dots, N)$ in a low dimensional space so that the dissimilarities $d_{ij}$ in the original state
are well-approximated by $|\mbox{\boldmath $u$}_{i}-\mbox{\boldmath $u$}_{j}|$, i.e. we find points $\mbox{\boldmath $u$}_i$ that minimize $\displaystyle \sum_{i,j} \left( d_{ij}-|\mbox{\boldmath $u$}_{i}-\mbox{\boldmath $u$}_{j}| \right)^2$.
In MDS, for the sample points $\mbox{\boldmath $s$}_i, (i=1,2, \dots, N)$ in the $m$-dimensional Euclidean space, we define the square distance matrix $D$ as $D_{ij} = | \mbox{\boldmath $s$}_i-\mbox{\boldmath $s$}_j|^2$, 
and introduce
\begin{eqnarray}
Z=-\frac{1}{2}JDJ=(JS)(SJ)^{\rm T}
\label{eq_hh03}
\end{eqnarray}
where  $S_{k\ell}$ is the $\ell$-th component of $\mbox{\boldmath $s$}_k$ and $\displaystyle J =(J_{k\ell}) = (\delta_{k\ell}-1/N)$ is called the centering matrix whose effect for $S$ is $(JS)_{k\ell}=S_{k\ell}-\sum_k S_{k\ell}/N$.
Next we decompose $Z$ into its eigenvalues and eigenvectors as $Z\mbox{\boldmath $p$}_i=\lambda_i \mbox{\boldmath $p$}_i$ with $\lambda_{i} \geq \lambda_{i+1}$.
Then we get $Z = P \Lambda P^{\rm T}$
where $\Lambda_{ij}=\lambda_i \delta_{ij}$ and $P_{ij}$ is the $j$-th component of $\mbox{\boldmath $p$}_i$.
Therefore, we obtain
the matrix 
\begin{eqnarray}
U = P\Lambda^{1/2}, \ (\Lambda^{1/2})_{ij}=\sqrt{\lambda_i} \delta_{ij}
\label{eq_hh04}
\end{eqnarray}
which corresponds to the matrix $JS$
and the new point ${\mbox{\boldmath $u$}}_i$ which is the $i$-th row vector of $U$.
Clearly the new points ${\mbox{\boldmath $u$}}_i$ are reconstructions of the original points $\mbox{\boldmath $s$}_i$ and recover the distance $|\mbox{\boldmath $s$}_i-\mbox{\boldmath $s$}_j|$.
If $\lambda_{n} \gg \lambda_{n+1}$, we can approximate $U$ by its projection into the subspace spanned by the eigenvectors $\{\mbox{\boldmath $p$}_1,\mbox{\boldmath $p$}_2,\dots,\mbox{\boldmath $p$}_n\}$. 

As we can find the points $\mbox{\boldmath $u$}_i$ only from the distances, we start with the geodesic distances $d_{ij}$ instead of $|\mbox{\boldmath $s$}_i-\mbox{\boldmath $s$}_j|$ and get the new low dimensional vector $\mbox{\boldmath $u$}_i$ corresponding to the $i$-th sample point on the filament.
Finally, we obtain $n$-dimensional representations $\mbox{\boldmath $u$}_i=(u_i^{(1)},u_i^{(2)},\dots,u_i^{(n)})$ whose distances
preserve the geodesic distances in the original space as much as possible. 
This procedure is essentially the same as the principal component analysis for the data matrix $U$ \cite{Tene00, William02}.


\subsection{Results for the
mass-spring model
}
\begin{figure}
\begin{center}
\includegraphics[width=7.5cm,clip]{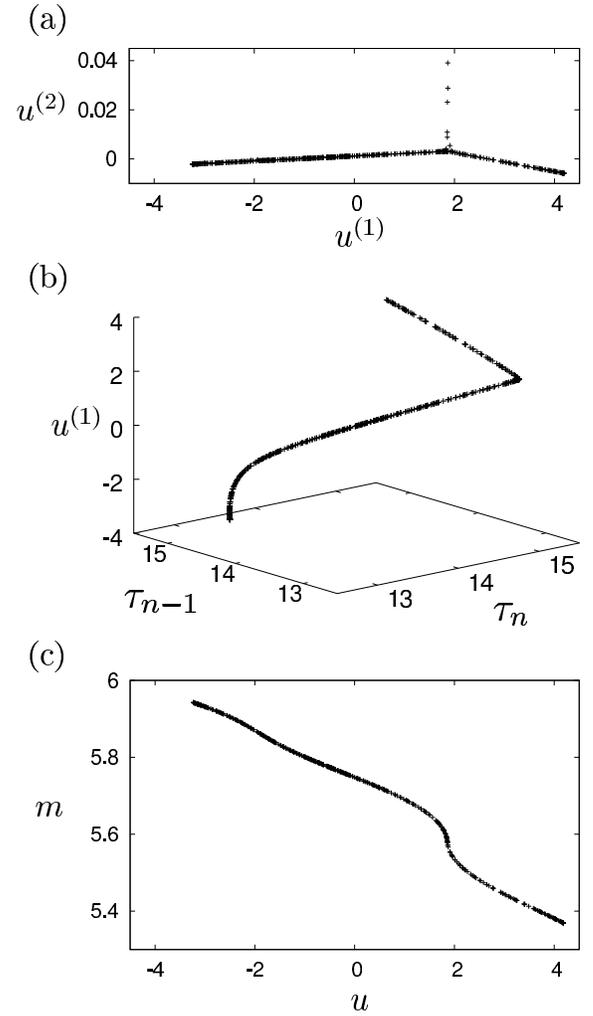}
\end{center}
\caption{
Results of the application of ISOMAP for Fig. \ref{f_hh2} (c). 
(a) Configurations of sample points onto the $\left(u^{(1)}, u^{(2)}\right)$ plane.
(b) Plot of sample points in the $\left(\tau_{n-1},\tau_{n},u^{(1)}_n\right)$ space and its projection onto the $(\tau_{n-1},\tau_{n})$ plane.
(c) Relation between the new coordinate $u$ and the droplet mass $m$.
}
\label{f_hh3a}
\end{figure}

Figure \ref{f_hh3a} (a) shows configurations of sample points for the dripping faucet data $(\tau_{n-1}, \tau_n)$ onto the $\left(u^{(1)}, u^{(2)}\right)$ plane obtained from ISOMAP.
Here, the spread of points in the direction of the $u^{(2)}$ component is much smaller than that of the $u^{(1)}$ component (about 0.6\%).
This means that the points on the filament are almost explained only by the first component $u^{(1)}$, which implies that ISOMAP succeeds to unfold the attractor to a straight line and $u^{(1)}$ is considered as a new coordinate along the filament (Fig. \ref{f_hh3a}(b)). 
Hereafter, $u_n^{(1)}$ is abbreviated to $u_n$ as we mainly use the first component.
The successful unfolding can be also confirmed from the one-to-one relation between the mass $m_n$ and the new coordinate $u_n$ in Fig. \ref{f_hh3a}(c).

\begin{figure}
\begin{center}
\includegraphics[width=8.5cm,clip]{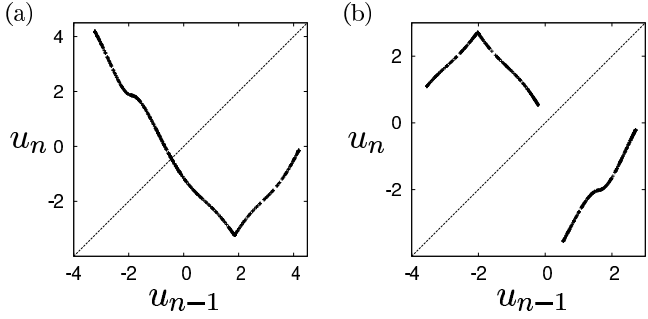}
\end{center}
\caption{
(a),(b)  Return plots of the new variable $u^{(1)}$ generated by ISOMAP for Fig. \ref{f_hh2} (b) and (c), respectively.
}
\label{f_hh3b}
\end{figure}

As any point on the filament is deterministically mapped to another point, the time evolution of $u$ is described by a one-dimensional map  $u_{n}= \varg(u_{n-1} )$ which is shown in Fig.\ref{f_hh3b}.
In addition, the points whose spread in the $u^{(2)}$ direction in Fig. \ref{f_hh3a}(a) is relatively large are located around the folding point (critical point) of the one-dimensional map in Fig. \ref{f_hh3b}(a).

The results in Fig.\ref{f_hh3b} show the expected relationship between $u_{n-1}$ and $u_{n}$.
We approximate this one-dimensional map $\varg(u)$ by a locally quadratic function
\begin{eqnarray}
\varg(u) = \sum_{j=0}^{2} a_j(u) u^j
\label{eq_hh1}
\end{eqnarray}
where $a_j(u)$ is determined by the locally least square method, i.e. $a_j(u)$ satisfies
\begin{eqnarray}
\min_{\{a_0(u),a_1(u),a_2(u)\}} \sum_{n}  (\varg(u_{n-1})-u_{n})^2 \exp\left\{-\left(\frac{u_{n-1}-u}{\sigma}\right)^2\right\}
\nonumber
\end{eqnarray}
and we use $\sigma=0.05$.
\subsection{Statistical properties}
From the one-dimensional map $\varg(u)$, we can get the natural invariant density $\rho(u)$
which is the base for the discussion of the statistical properties.
Here, $\rho(u)$ satisfies the equation
\begin{eqnarray}
\rho(u) = H(\rho(u)) &=& \int \rho(v) \delta(u-\varg(v)){\rm d} v
\nonumber \\
&=& \sum_{u_n:u=\varg(u_n)}
\frac{\rho(u_n)}{|\varg'(u_n)|}, \ \varg'(u)=\frac{{\rm d} \varg}{{\rm d} u}
\label{eq_hh2}
\end{eqnarray}
 where $H$ is called Frobenius-Perron operator and $\rho(u)$
is generally expected to be
the empirical distribution
$\lim_{N \to \infty} \sum_{n=1}^{N} \delta(u-u_n)/N $ for a chaotic orbit.

To solve approximately Eq.(\ref{eq_hh2}), we  divide the domain of $\varg(u)$ into the intervals 
${\rm I}_i, (i=1,2,\dots)$ whose edges are inverse mapping points $\varg^{-n}(u_*), (n=0,1,2,\dots)$, where $u_*$ is the critical point (the minimum in Fig.\ref{f_hh3b} (a) and the maximum in Fig.\ref{f_hh3b} (b) ) of the function $\varg(u)$ and expand $\rho(u)$ as
\begin{eqnarray}
\rho(u)=\sum_i \alpha_i e_i(u), \ \ e_i(u)=\begin{cases}
 1, (u \in {\rm I}_i) \\
0, ({\rm others}).
\end{cases}
\label{eq_hh3b}
\end{eqnarray}
We substitute Eq.(\ref{eq_hh3b}) in Eq.(\ref{eq_hh2}) and get 
\begin{eqnarray}
\alpha_i = \sum_{j} H_{ij} \alpha_j, \ H_{ij} = \frac{\beta_{ij}}{ |\varg'(u_j^{(\rm c)})|}, \ \beta_{ij}=\begin{cases}
1, \ {\rm if} \ {\rm I}_j \subset \varg({\rm I}_i)
\\
0, \ {\rm if} \ {\rm I}_j \nsubset \varg({\rm I}_i)
\end{cases}
\label{eq_hh2b}
\end{eqnarray}
where $u_j^{(\rm c)}$ is the center of ${\rm I}_j$.
The solution $\alpha_i$ is given by the eigenvector corresponding to the eigenvalue 1 of the matrix $H$.
The above-mentioned method is a kind of the Galerkin-approximation which is often used in the study of the one-dimensional map \cite{Kohda90}.

\begin{figure}
\begin{center}
\includegraphics[width=8.5cm,clip]{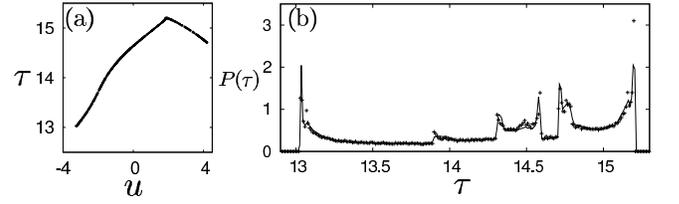}
\end{center}
\caption{(a) Relation between the new valuable $u$ and the dripping interval $\tau$. (b) Distribution function of dripping intervals $P(\tau)$ which is derived from Eq.(\ref{eq_hh4}). 
The result from the direct simulation ($10^6$ droplets) are also plotted by the symbol +.}
\label{f_hh4}
\end{figure}

The internal state variable $u$ is not a natural physical quantity for the dripping faucet system.
However we can derive any physical quantity $A$ from $u$, because
the quantity $A$ on the filament is determined by $u$, i.e., $A=A(u)$ and its long-time average is calculated by $\left< A \right>=\int A(u)\rho(u) {\rm d} u$.
Actually, as the results of ISOMAP provide the relationship between $u_n$ and $(\tau_{n-1},\tau_n)$ as shown in Fig.\ref{f_hh4}(a), the important observable variable $\tau$ of the dripping faucet system is determined by 
\begin{eqnarray}
\tau=\phi(u)
\label{eq_hh31}
\end{eqnarray}
where the function $\phi(u)$ is approximated in the same way as $\varg(u)$ (see Eq.(\ref{eq_hh1})). 
First we get the distribution function
\begin{eqnarray}
P(\tau)=\frac{\rho(u)}{\left| \phi'(u) \right|}
\label{eq_hh4}
\end{eqnarray}
which is one of the most basic properties of the dripping faucet.
The result in Fig. \ref{f_hh4} shows $P(\tau)$ with many peaks which are generated by the folding processes of $\varg(u)$.
These properties are consistent with the result from the direct simulation of Eq.(\ref{eq2-1c}) and (\ref{eq2-2}).
\\

Next, we calculate the Lyapunov exponent \cite{Ott02} and the topological entropy \cite{Hao98} which characterize the stability and the variety or complexity of chaotic orbits, respectively.
The Lyapunov exponent of $\varg(u)$ is defined by
$
\Lambda = \lim_{N \to \infty} (1/N) \ln \left| {\rm d}u_{N}/{\rm d}u_0 \right|
=\lim_{N \to \infty} (1/N) \sum_{n=0}^{N-1}  \ln \left| {\rm d}u_{n+1}/{\rm d}u_n \right|
$
and given by
\begin{eqnarray}
\Lambda =\left<\ln \left| \varg'(u) \right| \right>.
\end{eqnarray}
The topological entropy is equal to the largest eigenvalue of the transfer matrix $\beta$ in Eq.(\ref{eq_hh2b}).
As both quantities are invariant under the transformation from $u$ to $m$,
the results from $\varg(u)$ should coincide with the Lyapunov exponent and topological entropy of $f(m)$ 
(if the map $\varg(u)$ is an appropriate description of the dripping faucet dynamics given by Eqs. (\ref{eq2-1c}) and (\ref{eq2-2})).
They are cited in Table \ref{t_hh1}.
The Lyapunov exponents are in good agreement,
but the topological entropy from $\varg(u)$ is slightly smaller than the results from $f(m)$.
This may mean that the sample points $\tau_n$ do not include rare orbits, because
the results are based on only 500 sample points.

\begin{table}
\begin{center}
\caption{Lyapunov exponents and topological entropies calculated from $f(m)$ and $\varg(u)$.
}
\label{t_hh1}
\begin{tabular}{lccc}
\hline
\multicolumn{1}{c}{$v_0$} & \multicolumn{1}{c}{$f(m)$ or $\varg(u)$} & \multicolumn{1}{c}{Lyapunov Exp.} & \multicolumn{1}{c}{topologcal entropy} \\
\hline
 0.1128 & $f(m)$ & 0.253 & 0.859\\
       & $\varg(u)$ & 0.253 & 0.855\\
\hline
 0.1129 & $f(m)$ & 0.306 & 0.913\\
       & $\varg(u)$ & 0.291 & 0.882\\
\hline
 0.1130 & $f(m)$ & 0.350 & 0.948\\
       & $\varg(u)$ & 0.346 & 0.903\\
\hline
\end{tabular}
\end{center}
\end{table}
\begin{figure}
\begin{center}
\includegraphics[width=8.5cm,clip]{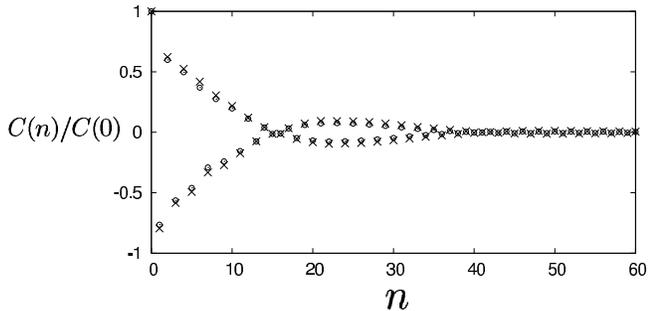}
\caption{
Time correlation function of $\tau_n$. The results from $\varg(u)$ ($\circ$) and the direct simulation ($10^6$ droplets) from Eq. (\ref{eq2-1c}) and (\ref{eq2-2})($\times$) are plotted.
The two curves show the decay for a period-2 and a long period (about 23) oscillation.
}
\label{f_hh5}
\end{center}
\end{figure}
We can calculate the time series $\tau_n$ from the one-dimensional map (\ref{eq_hh1}) and Eq.(\ref{eq_hh31}), but its long time behavior has a large difference from the direct simulation of Eq.(\ref{eq2-1c}) and (\ref{eq2-2}), because the dynamics is chaos.
We also calculate the time correlation function of $\tau_n$
\begin{eqnarray}
C(n)=\lim_{N \to \infty} \frac{1}{N}\sum_{m=0}^{N-1}(\tau_m-\bar{\tau})(\tau_{m+n}-\bar{\tau})
\nonumber \\
=\left< \phi(u) \phi(\varg^n(u)) \right>-\left< \phi(u)^2 \right>
\label{eq_hh5}
\end{eqnarray}
where $\bar{\tau}=\lim_{N \to \infty} (1/N)\sum_{m=0}^{N-1}\tau_m$.
The result in Fig.\ref{f_hh5} shows the exponential decay for a period-2 and long period (about 23) oscillation.
Their properties are also shown in the result which is calculated directly from $\tau_n$ by the simulation of Eq. (\ref{eq2-1c}) and (\ref{eq2-2}).
The good coincidence shows
that by using the internal variable $u$, we can discuss not only the static properties but also the dynamical property of the original observable $\tau$.
\subsection{Application to real experimental data}
\begin{figure}
\begin{center}
\includegraphics[width=7cm,clip]{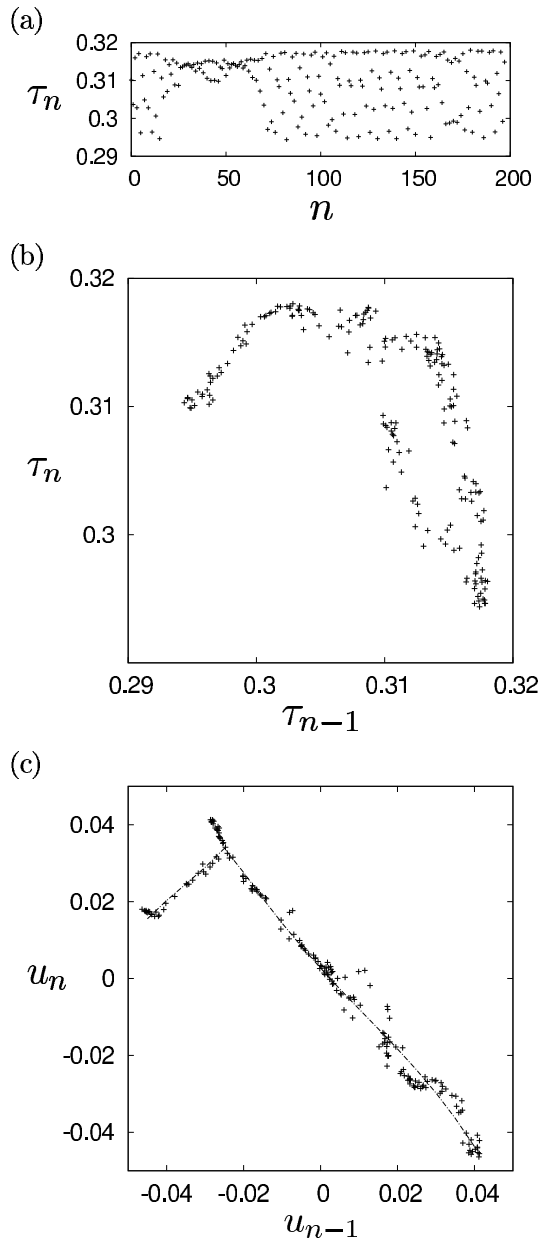}
\caption{
Results for our experimental data.
(a) Dripping time series $\tau_n$ which contains one or two periodic motion around $n=50,170$.
(b) First-return plot of $\tau$. It shows a multi-valued function in the right region. 
(c) First-return plot of inner state variable $u$ which suggests that there is a one-dimensional map $\varg(u)$. It may has a wavelike pattern in the right region.
}
\label{f_hh6}
\end{center}
\end{figure}

Last we show preliminary results of the application of the dimension reduction method to our real experimental data. 
A short time series whose water flux is almost stationary 
and its first-return plot $\tau_{n-1}$ vs. $\tau_n$ are shown in Figs. \ref{f_hh6}(a) and (b), respectively.
This return plot shows a multi-valued function in the wide region and cannot describe the dripping faucet dynamics. 
The application of our method
leads to the internal variable $u$ and the one-dimensional map $u_{n}=\varg(u_{n-1})$ which is shown in Fig. \ref{f_hh6}(c).
The absolute value of its slope is nearly one in almost all regions i.e. the instability of the orbits is weak which is related to the fact that the time series contains one or two periodic like motions.

It has been pointed out
that ISOMAP is topologically unstable for small noise \cite{Balasubramanian02}.
In actually, the neighboring graph $\cal G$ around the folding point of the filament is affected by experimental noise or high-dimensional dynamics and has some short cuts which are out of the filament.
Therefore, the first return map of $u$ has a muliti-valued structure around the critical point (maximum point) of $\varg(u)$.
However, this effect is small and localized so we can say our method is a promising method for not only the numerical study but also experimental data.

Our experiment is a first step and currently we cannot keep it stationary to measure long time series.
We work on a revised set-up and we shall present extended experimental results in a future article.

\section{Summary and Discussion}
In this paper, employing the dripping faucet system as an illustrative example, we studied the problem of constructing a surrogate variable for the internal state of the chaotic dynamical systems from time series using manifold learning analysis.  
Especially, when the time-delay embedding of the observed time series forms a one-dimensional curved structure,  we succeeded to obtain one-dimensional deterministic maps associated with the surrogate variable.
The statistical properties of the original chaotic system were successfully reproduced by its surrogate system. 
 
In real-world applications, not all original state variables of the system can be directly observed.
Instead, only some of the original state variables or their transformations are observed in experiments.  
So, it is often seen that the manifolds obtained from the time-delay embedding may be very complex even if their dimensionality is low, which leads to multi-valuedness in the return plots~\cite{Chr97}. 
For example, the time series of inter-spike intervals is mainly observed in neural systems and its return plot often exhibits a multi-valued function~\cite{Feud00}. 
Besides neural systems, there is a number of such examples, e.g., laser systems~\cite{Hubn93,Used09}, passive biped walkers~\cite{Gosw98}, and social activity models~\cite{Feic95}.
Extracting the deterministic relationship from the observations of these models may be also done using dimension reduction methods.  
 
Besides, the return plot  of $u_{n-1}$ vs. $u_n$ (Fig. \ref{f_hh3b}) is less smooth compare to that of $m_{n-1}$ vs. $m_n$ (Fig.~\ref{f_s3} (a)).
This is because methods of manifold learning are generally unsupervised ones, just using the information of $\tau_n$. 
If the assumption that the training data is generated from a dynamical system with a simple mapping form say the logistic parabola can be incorporated additionally to the manifold learning as some constraint or penalty terms, we can obtain a more refined  return plot which may be more interpretable to us.   
We would like to develop such problems in our future works.

In this paper, we have been only concerned with the case in which the internal state behind the observed time series obeys a one-dimensional dynamical system,  this is of course an ideal case.
We have to extend our approach to higher dimensional cases (two, three dimensional maps). 
As formal methodologies (application of ISOMAP or other manifold learning methods) are not restricted to one dimensional manifolds, the presented approach can in principle be extended to higher dimensional cases.
It should be noted, however, that ways of acquiring training samples to obtain a lower dimensional representation become more important.
For example, let us consider the situation where  the H\'{e}non attractor ${\cal A}$ is lying on a two dimensional nonlinear manifold ${\cal M}$ embedded in, say, the three dimensional Euclidean space ${\mathbb R}^3$.
In order to obtain a lower dimensional representation of ${\cal M}$, not only the data on ${\cal A}$, but also the data associated with transient dynamics are needed because ${\cal A}$ is too thin to recover the whole two dimensional structure of  ${\cal M}$.    
In addition, the non-uniformity in the natural measure on ${\cal A}$ and its transient area affects the performance of manifold learning.

\begin{acknowledgments}
This study is partially supported by Grant-in-Aid for Scientific Research (No. 22740258), the Ministry of Education, Science, Sports, and Culture of Japan.
The research leading to the results has received funding from the European Community's Seventh Framework Programme FP7/2007-2013 under grant agreement No. HEALTH-F2-2009-241526, EUTrigTreat. 
Furthermore, support by the Bernstein Center for Computational Neuroscience II G\"ottingen (BCCN grant 01GQ1005A, project D1) is acknowleged.
H.S. is grateful to S. Akaho for fruitful discussions and comments. 

\end{acknowledgments}

\begin{thebibliography}{4}

\bibitem{Shaw84}
R. Shaw, {\em The Dripping Faucet as a Model of Chaotic System} (Aerial Press, Santa Cruz, 1984). 

\bibitem{Kiyo99}
K. Kiyono and N. Fuchikami, J. Phys. Soc. Jpn. 68, 3259 (1999).

\bibitem{Fuch99}
N. Fuchikami, S. Ishioka, and K. Kiyono, J. Phys. Soc. Jpn. {\bf 68},  1185 (1999).

\bibitem{Kats99}
T. Katsuyama and K. Nagata, J. Phys. Soc. Jpn. {\bf 68},  396 (1999).

\bibitem{Chr97}
F. Christiansen, P. Cvitanovi\'c and V. Putkaradze, Nonlinearity {\bf 10},  55 (1997).



\bibitem{HTF09}
T. Hastie,  R. Tibshirani, J. Friedman, {\it  The Elements of
Statistical Learning: Data Mining, Inference, and Prediction}
(Springer-Velag, New York, 2001).

\bibitem{Tene00}
J.B.~Tenenbaum, V.~de Silva and J.C.~Langford, 
Science {\bf 290}, 2319 (2000). 

\bibitem{Cox00}
T.F.~Cox and M.A.A.~Cox, {\it  Multidimensional Scaling}
(Chapman and Hall, London, 2000).











\bibitem{William02}
C. Williams,
Machine Learning {\bf 46}
11 (2002). 
\bibitem{Kohda90}
T. Kohda and K. Murao,
IEICE {\bf E73} 793(1990).
Erik Bollt, Pawe\l\ \'Gora, Andrzej Ostruszka and Karol \.Zyczkowski, 
SIAM J. on Applied Dynamical Systems, {\bf 7}, 341(2008).


\bibitem{Ott02}
E. Ott, 
{\em Chaos in Dynamical Systems} (Cambridge University Press, Cambridge, 2002).

\bibitem{Hao98}
B.-L. Hao, W.-M. Zheng
{\em Applied Symbolic Dynamics and Chaos}(World Scientific Publishing Company, Singapore, 1998).

\bibitem{Balasubramanian02}
M. Balasubramanian and E. L. Schwartz,
Science {\bf 295}, 7, (2002).

\bibitem{Feud00}
U.~Feudel, A.~Neiman, X.~Pei, W.~Wojtenek, H.~Braun, M.~Huber and F.~Moss, 
Chaos {\bf 10},  231 (2000).


\bibitem{Hubn93}
U.~H\"{u}bner, C.-O.~Weiss, N.B.~Abraham and D.~Tang, 
In: A.S.~Weigend and N.A.~Gershenfeld (eds.), {\em Time-Series Prediction: Forecasting the Future and Understanding the Past}, 73, (Westview Press, Boulder, 1993). 

\bibitem{Used09}
J.~Used and J.C.~Mart\'{i}n, Phys. Rev. E {\bf 79}, 046213 (2009); {\em ibid.} {\bf 82}, 016218 (2010). 

\bibitem{Gosw98}
A.~Goswami, B.~Thuilot and B.~Espiau, 
Int. J. of Robotics Res. {\bf 17}, 1282  (1998). 

\bibitem{Feic95}
G.~Feichtinger, L.L.~Ghezzi and C.~Piccardi, 
Int. J. Bifurcation and Chaos {\bf 5}, 255  (1995). 

\end{thebibliography}

\end{document}